\documentclass[CRMECA,Unicode,manuscript]{cedram}

\usepackage[all]{xypic}

\title[]{Ramblings (Memoirs) of a Scientist}

\author{\firstname{Jay} \lastname{Fineberg}}
\address{The Racah Institute of Physics, The Hebrew University of Jerusalem, Givat Ram, Jerusalem 91904, Israel
}
\email{jay@mail.huji.ac.il}
\thanks{The  author is supported by Israel Science Foundation through grant \#840/19
  } 



\begin{abstract} 
This paper is a faithful description of the author's career as a scientist, which often intersected that of Yves Couder. The emphasis of this paper is a true description of how the science that the author has been associated with really came about. Included are  brief descriptions of the science associated with the research paths described.  It is hoped that this rather accurate account may be amusing for the senior scientists among us and educational (and possibly useful) for younger scientists.  

\end{abstract}

\begin{document}

\maketitle


\selectlanguage{english}

\section{Introduction}

Yves Couder was a good friend and an amazing scientist. His death was a terrible loss to both science and to me, personally. It is extremely rare to find scientists such as Yves. His imagination, ingenuity and academic honesty were an inspiration to all of us who had the incredibly good fortune to have known him. 

I was asked to contribute a paper in Yves’ memory. I, of course, immediately agreed to do it. In the next instant, I immediately regretted agreeing – but my finger had already pressed ‘send’. In this age of instant gratification, often a random impulse entails an unwanted commitment. I should say that, in general, I don’t enjoy writing papers (I do have a few to my name, but each has been a labor of love/angst/frustration that is comparable to giving birth to a child, which is easily said, since I never had to perform that particular task).  I do force myself to write papers – but only if I really have something that I believe is worth saying. As such, this particular commitment was especially challenging. I didn’t want to submit an ‘ordinary’ paper – one that summarized, in general,  a partial answer to (hopefully) a good scientific question. 
 
After some deliberation and procrastination (both crucial pieces of the scientific process), I decided to write down a true and unembellished account of how I arrived  (scientifically) at where I am now. The reasons for this decision are that (1) I believe that Yves would have been amused by this (2) I don’t believe that young scientists are generally exposed to such things (at least not publicly). I would like to credit any good that comes out of this endeavor to Yves and take upon myself any of the negative results. Why do I believe that these personal revelations may be important to my younger colleagues? Generally, every time that we publish an account of our work in a scientific journal we ‘rewrite’ history. Our results are always set out in a clear and logical order in which we first introduce a concise scientific question and then go on to try to convince our readers that we have, to some degree, answered it. This logical and well-ordered sequence of events is, from my personal experience, the antithesis of the true scientific process.   Generally, the direction that we choose when embarking on any new scientific research is based on entirely the wrong initial question (e.g. Columbus’s voyage to discover a passage to India). Only after years of banging our collective heads against the wall do we find true (or partial) enlightenment; our results have steadfastly dragged us, kicking and screaming, to the realization that we have been asking the wrong question the whole time. This process is never, of course, noted in the scholarly paper that comes out of this work. Somehow, we always neglect to mention that, for 95$\%$ of the seemingly infinite period that was spent performing this study, we were entirely misguided and (in retrospect) were total idiots. 

I start with a few words about myself. I’m an experimental physicist who comes from a long line of contrarians. Perhaps this characteristic is a good one to have for a scientist, as it is never a good idea to entirely believe what is generally assumed to be true. In science, what is often conceived to be correct today may well be overturned tomorrow. My personal bias is that unwavering belief is a trait better retained for religion. Science is the art of doubt. Being a natural (genetically disposed) contrarian (in contrast to a contrarian forged by circumstance) perhaps affords a certain advantage in science. This trait also extends to my own research; I doubt my own discoveries almost as much as I doubt the results of others. A good scientist, in my opinion, is always afraid that someone will repeat his/her work and find out that he/she missed something critical. Personally, I’m always surprised when someone else’s work actually supports my own. In my experience, it is always important to evaluate and reevaluate our preconceptions. If possible, it is always worth going back to the raw data – just to make sure that “in the light of day” it still supports the scientific house of cards that we have constructed out of it. 

Other character flaws that I possess are an innate hate of school (I have always looked for the back door, when pushed into the front one in the various educational institutions that I have attended) and having been, for most of my life, slightly nearsighted. This first trait may well be due to my contrarian nature; the fact that I am still in school could be seen by those who know me as a sort of a Karmic revenge, although the jury is still out on this one. The second trait has actually significantly shaped my character, in a sense. Having a fuzzy, rather myopic, view of the world around me has served to push me into continually looking for the “big picture” (as the various little pictures have never seemed to be in focus.). I believe that a constant search for the “big picture” is a good one for a scientist. So long as we are spending time beating our heads against a wall, the wall might as well be a big one. Of course, I’m not claiming that, in order to see the big picture, we should have bad vision – but it sometimes is useful to really {\em not} see the trees in order to see the forest.

\section{The beginning – the formative years}
I was always interested in how things fit together. Indeed, I spent quite a lot of time during my formative years destroying various things in my parents’ basement and, generally unsuccessfully, trying to get them to fit back together. It was then obvious (at least to me) that my field of study should be Physics (while throwing in a degree in Mathematics, for fun). I started this endeavor after a long hiatus in which I served in the Israeli army for a few years, followed by a stint in a kibbutz in which I successfully picked apples and much less successfully tried my luck with female volunteers. By the end of my studies I had become successfully brainwashed by my professors (despite my contrarian nature) to believe that true physics can only be done by brilliant theorists and that the best of the best belonged to a field then known as particle physics and now known as high energy physics. We were carefully taught that every problem had an answer, generally an analytical one. Moreover, all that one needed to do to solve any given problem was to think deeply and play around with various strange functions and the answer would come. The proof of this was in every exercise sheet that we were required to solve; each had a clear and well-defined answer that was generally composed of various analytic functions.

Although I performed these tasks with aplomb, I started getting the feeling that I was missing something. Maybe the feeling resulted from the numerous apples that had been falling on my head throughout my kibbutz life (for Newton it, apparently, only took one apple – but not all of us are Newton). The more that I learned, and the better my grade average got, the less I felt that I understood. I completed my degrees with both excellence and a feeling of total inadequacy. This feeling continues to this day, although I have the suspicion that I have quite gotten used to it. 

At any rate, I started my graduate career as a theorist in (you guessed it) high energy physics.  Formally, I was in a 2 year struggle to calculate the cross section for the formation of magnetic monopoles by interacting photons. Informally, I was in a two year struggle to get my motorcycle, an old (now “classic”) Triumph Tiger to keep running. The latter pursuit took up most of my time, whereas the former paid for the repair bills. At the end of those two years I did uncover  \cite{Fineberg_monopoles_1985} that the monopole cross-section scaled as $e^{-\pi/\alpha} \sim  e^{-137\pi}$, where $\alpha$ is the fine structure constant. This impressive result, which has since been cited twice to my knowledge (by citing this paper here, I have increased its citations by 50\%) had a major effect on my scientific career. Proudly flourishing my result, I immediately asked my elders if I had gotten the answer right. The response that I received was less than heartwarming, “well, since we don’t see any monopoles around, your result is certainly consistent”. This faint praise was not quite the satisfaction that I was anticipating, when I had made the bold decision to spend two years (formally, of course) in solving intractable integrals. I decided that, if this is the satisfaction that theory provides, then I may as well go back to the kibbutz and retry my luck with the women volunteers. It was also then that I had a small epiphany. I hadn’t entirely wasted these formative years in physics, my motorcycle was still running (on average)!  I had developed an entirely unforeseen ability to put things back together so they would (on average) work. Armed with this important realization, I decided both to continue on to a PhD and to “cross the lines” from theoretical to experimental physics. 
 
At the time, the decision to go from the “cutting edge” of theoretical physics to experimental physics was not a standard one. People in my department were fairly surprised, and a few eyebrows were raised. I, however, realized that by doing experiments I would always get a “correct” answer.  I did not realize, at the time, that the art in Physics was not necessarily to get an answer, but essentially to figure out the question that I should have asked. I also have realized over the years, that often the question is actually more interesting (and important) than the answer.

\section{Onward to a Ph.D.}	
I started my career in experimental physics with Victor Steinberg, then a new faculty member who had just started his laboratory. Victor is an amazing scientist who started his own career as a (Russian) theorist. When he couldn’t find anyone to test his theories, he decided to become an experimentalist. To this end, he spent a few years in Guenter Ahlers’ lab at UCSB to learn the trade \cite{Ahlers_Victor_1985}.

  Guenter had been investigating chaotic systems in convecting He3-He4 fluid mixtures \cite{Ahlers_berhringer_1978}. The idea of chaos was, at that time, entirely new and was on its way to eventually becoming the new field of nonlinear physics. Guenter, together with Victor, had developed the first visualization methods to be able to directly observe what was really taking place in these intriguing types of systems \cite{Ahlers_Victor_1985}.  When I first met Victor, he had just arrived at the Weizmann Institute of Science. I found him sitting in a dirty and dusty lab that was full of someone else’s old junk. He had neither students nor experimental equipment but he was holding a pile of fuzzy and rather unintelligible pictures of something, and was obviously quite excited about it. He immediately set forth on a long disposition in heavily accented Russian English (or maybe it was Hebrew – or quite possibly both). After listening patiently, I realized that I didn’t understand a single word about what he was trying to convey. On the other hand, the pictures looked kind of interesting, so, on the spot I told him that I would be glad to do a Ph.D. in his lab.  I then embarked on a 4 year journey to try to figure out what Victor had been trying to tell me. 

At this point in these ramblings I have come to a crossroads. My intention in writing this is to attempt to pass on my personal history to younger scientists in a, probably misguided, attempt to present a non-standard perspective about how science evolves. I did not intend to seriously describe the science, itself, that I have actually been involved in. On the other hand, my science and personal history are rather entwined. I believe that, at least a general description of the science is needed to provide context. I will, therefore, try to present as brief a description of the science as possible, without going into any detail. If anyone is interested in the details or science, I’ll reference some of the relevant work. 

Although I had no idea of this, when I decided to delve into trying to understand Victor’s pictures, the first ideas of chaos and driven nonlinear systems were just starting to unfold.  The idea of chaos, in particular, had taken scientists pretty much by surprise. The fact that entirely deterministic rules could lead to seemingly random, unpredictable behavior was something of a revolution in scientific thought. These systems were, furthermore, classical systems where Newton’s laws rigidly applied. Logically, we should have been able to predict everything about the future evolution of these physical systems, but we found that often we had nearly zero predictive ability. These ideas defied the `common knowledge' at the time. This was an exciting period in which more and more physicists joined the battle; to both characterize these systems and understand their predictability. 

 At this time, no one really contemplated the role of the spatial degrees of freedom in such systems. The behavior of nonlinear systems was mostly studied, experimentally and theoretically, within the time domain, essentially as kind of crazy nonlinear oscillators.  The language that had been developed to describe equilibrium phase transitions was found to be incredibly useful in these driven, far from equilibrium systems. People started to understand chaotic dynamics in terms of a control parameter (analogous to, say, temperature in equilibrium system) that described the distance from a phase transition (e.g. liquid to solid transition) and an “order” parameter that described the system’s response to changes in the control parameter.

The intellectual environment during these years was incredibly rich and exciting. Once the language of phase transitions was adopted, we could look at a huge range of, on the face of it, diverse physical systems in the same context – and with the same language. People realized that one could consider the simplest possible system to manipulate and understand the behavior of a broad class of infinitely more complex systems (the beauty of universal behavior). Conferences in this growing field typically discussed an extensive variety of entirely different physical systems, ranging from mechanical oscillators to diverse fluid flows to chemical reaction dynamics. The amazing thing about these gatherings was that the whole gathering had a common language and common intellectual framework. Each participant could follow the developments in any of these diverse fields. Yves had also grown up in this community, arriving after starting as a solid state physicist. The conference in honor of Yves’ 70th birthday was one of the last of these types of gatherings, that I am aware of. The extent of the scientific fields represented there was impressive and reflected different sides of Yves’ broad career. All of the older participants could follow everything, as we all grew up in this culture. I’m not sure whether the younger participants, who did not have the benefit of this broad education, benefited as much from it. 

My PhD project was in the study of thermal convection. These experiments were among the first in physics that were both visualized with extremely high sensitivity and every aspect of both the control and data acquisition in these experiments (a typical experiment studying convective patterns often required months to develop) was computer-controlled. 

 The first PCs had just been invented and every aspect of their use needed to be developed. For example, we could only visualize the flow patterns if a camera could both somehow be interfaced to a PC and, moreover, the data could be displayed on the computer screen and/or dot-matrix printer.  This last sounds a bit trivial, but at the time, no drivers existed to display anything but text on computer monitors, much less to print these images out. To this end, my lab mate, Elisha Moses, and myself were forced to teach ourselves both the requisite electronics as well as the sorely needed computer skills (i.e. eventually, writing assembler routines to draw images on the screen and printer – this only after figuring out how the screen memory and dot-matrix printer input were mapped). This turned out to be time well-spent, as we found ourselves at the technological forefront of a number of new fields. These experiments could not have been performed otherwise. 
 
   This is also a good example of when new technology can give you a real scientific edge. Visualization methods had been around for years, but the ability to conduct precisely controlled experiments over long periods of time (months), while continuously collecting all of the requisite data provided us with the capability to perform experiments that could not have been conceived of previously.  
 
Thermal convection is one of the prototypical systems that were used within nonlinear physics to understand first chaos and, later, to develop theories of pattern-forming nonlinear systems. The first project that Victor suggested was to study the propagation of a nonlinear pattern-forming front, separating conductive to thermally convecting regions of a system \cite{Fineberg_vortexfront_1987}. A theory had been suggested \cite{Dee_Langer_1983} in which such fronts were predicted to both exist and propagate at a well-defined dimensionless velocity of $2\epsilon ^{1/2}$, where $\epsilon$ is the system’s control parameter. In the convective system, $\epsilon$ was defined as the dimensionless temperature difference, $\epsilon \equiv (\Delta T- \Delta T_c))/\Delta T_c$ where $\Delta T$ is the temperature difference imposed in the system and $\Delta T_c$ the critical temperature difference at the onset of convective motion. This is where I learned to perform precise and controlled experiments. As we needed to verify both the scaling law as well as the coefficient, $\epsilon$ was varied over the wide range of $10^{-4} - 10^{-1}$. Extreme attention to detail was required to achieve the necessary 0.1mK$^\circ $ temperature stability for this type of room-temperature experiment. Visualization of the flow patterns that were excited was obviously a necessary requirement, to be able to follow the spatial and temporal behavior of these convection fronts. 

Many of the necessary skills for running such a precise and highly controlled experiment were taught to me (often accompanied by kicking and screaming, by both mentor and mentee) by Victor. I’m very grateful (in retrospect) for this strict (Russian, with some German traits inherited from Guenter) training, as I learned both that things like this can be done and how to do it. I also realized, like many other weapons in one’s arsenal, that you should push yourself to achieve this type of precision only when it is needed. Ultra-precise experiments are a means to achieve a goal, and not the goal itself. I also learned the rule of thumb of doing the best that you can in precision, as long as it is not too painful. Why? Because you never know when that extra bit of precision will teach you something that you might have missed otherwise. 

 The experiment was very successful; we indeed verified the theoretical result to a few percent accuracy \cite{Fineberg_vortexfront_1987}. I believe that this result has found its way into textbooks etc. The project, however, was not quite as successful as the experiment for the following (very surprising) reason. We, basically, did too good a job. There was a well-defined question – and we experimentally showed that the theory that was purported to answer the question was perfectly correct. As a result, there were no follow-up experiments needed (and, therefore, relatively few citations of this work at that time). More importantly, since we had closed the question, I had to find something else to do in order to complete my Ph.D. thesis (one paper just didn’t cut it in those days).
 
While at the time this produced a bit of consternation for the young PhD student (me), who basically wanted to complete his thesis work in less than his entire lifetime, this actually turned  out to be a stroke of luck for me. My lab-mate, Elisha Moses, had been given, in retrospect, a much `better’ project by Victor. Elisha, who was my good friend, later my brother-in-law and, to this day, my good friend {\em and} brother-in-law (neither of those last statements are at all trivial), was given the task of looking at convective flow in cognac (or water-alcohol ‘binary fluid’ mixtures).  The idea was to verify earlier theoretical work of Victor’s that had predicted that thermal convection under the right conditions (the alcohol/water percentage of good cognac) would lose stability to temporal oscillations. I suppose this is not unlike the behavior of a person that has imbibed too much cognac, but for different physical reasons.

 This question could have ended up like my own first project, by simply verifying the theory. Fortunately, Elisha and Victor immediately encountered entirely unexpected behavior. This is an example of some of the very best theories, in my view. The best theories are those that are wrong, although not {\em entirely} wrong. The best theories should be correct enough to both provide some intuition and to cast you into an entirely unforeseen direction that is more interesting and, ultimately important, than what you were interested in at first. 

What happened when the experiment was initiated? Well, Victor’s theory wasn’t entirely off. At approximately the correct region  of phase space (here, the alcohol concentration), oscillating fluid flow at each spatial point was indeed observed.  The behavior in space and time of this system was, however, quite a bit ‘richer’ than anyone had remotely expected \cite{Moses_multistability_1987}. (For the uninitiated, ‘richer’ in science means, in general, that the observed behavior is initially totally incomprehensible and appears impossibly complicated. The term ‘richer’ is generally used in the literature when, in retrospect, researchers have started to make some sense of the system. Note, however, that scientific articles are always written in retrospect. You can generally be sure that in a ‘rich’ system a lot of beating of heads on the wall preceded all of this ‘richness’.)  In our case, the richness of the system initially presented itself as an uncontrollable and seemingly senseless ménage of randomly appearing and disappearing convecting patterns, The form and size of these patterns seemingly defied intelligent description.

 When Elisha and Victor first observed this mess, they spent a considerable time (as good experimentalists should) looking for random leaks in the system, bugs in the software, or Maxwell demons and other kinds of leprechauns who were playing havoc at their expense. After running through a gauntlet of sanity checks, they eventually convinced themselves that this very weird behavior was indeed entirely real; they had uncovered a treasure trove of new and unexpected physics. They came to this realization just in time, by the way. At the same time there were 3 other experimental groups who were conducting precisely the same type of experiments in nearly the same region of parameter space. All of these groups were extremely good, well-known, and better equipped than we were. The resulting explosion of weird results in this system was, eventually, good for everyone involved. This concentrated activity generated a flurry of frenzied scientific (theoretical and experimental) activity dedicated to the study of this unexpected behavior within an archetypal nonlinear system. 
 
 Yours truly was thrown into this scientific fray. This turned out to be a decisive advantage in this competitive scramble for scientific enlightenment, since Elisha and I had two parallel and independent working experiments that could both contribute. We were quite successful, actually. Together and separately (with Victor’s guidance) we came up with a number of key results that contributed to formulating ideas, validating others and pushing the theorists as well as our competitors to come up with new ways to make sense of the fascinating new phenomena that we all observed \cite{Fineberg_Moses_PRL_1988}.  
 
What did this work accomplish at the end of the day? I believe, in retrospect, that the major impact of this work was to demonstrate that nonlinear interactions were not confined to ‘simple’ chaotic systems whose behavior is solely temporal. These spatially extended nonlinear systems demonstrated complexity (and ‘richness’) that stemmed from the fact that propagative effects are important. Nonlinear fronts could and did propagate throughout nonlinear systems and generate local nonlinear behavior (e.g. nonlinear patterns) that was not necessarily correlated to what was taking place elsewhere. These experiments also revealed the importance and prevalence of nonlinear self-focusing behavior that could generate localized nonlinear states. The measure of the importance of this work is always in retrospect; much of this work is still being routinely cited.

{\em Science aside, what did I learn from my Ph.D.? }
\begin{itemize}
\item {\em Technology is useful.} If you can find a new way of looking at a physical phenomenon that could not have been technically achieved previously, you will generally be surprised. While sometimes a ‘surprise’ will simply show you that you are an idiot – and should have known better, surprises in science are often good things. The successful surprises are often (in retrospect) redefined as ‘discoveries’.  While the ‘unsuccessful’ surprises generally don’t see the light of day in scientific publications, they also have an important purpose. So long as we learn from them, we may not turn out to be the same kind of idiots in future work. 
\item {\em Independence should be nurtured.} In the last year of my Ph.D., Victor went on a Sabbatical and left me on my own for over a year. In those days, the internet was only beginning. As a result, communication was relatively poor and, generally, back and forth messages were delivered entirely out of phase with one another. It was only during this stage that I developed scientific independence. Victor is a brilliant researcher, who generally had answers to any and all issues before I, a student, could even formulate the questions. As a result, until I was left alone, I was more a technician than a scientist. Only when left to my own devices, did I start to think independently and critically.  It was during this (painful, at first) period that most of the papers and insights that resulted from my doctorate came about. This is not to say that Victor didn’t have a major influence on these results. Once I managed to come up with and verify an idea, our mutual discussions were critical in figuring out what these findings meant and where they fit into (or didn’t fit into) the overall scientific picture.
\item{\em Head-banging as part of the process.}  Banging one’s head against the wall is an integral part of the scientific  and creative process. Head banging is a painful but entirely necessary. How else is the wall going to move (or alternatively, are we to figure out that maybe it is best to circumvent it?)? The freedom to do this (i.e. the crux of scientific independence) is an important thing for both students and advisor to realize. It is critical that we (advisors) leave our students space to develop their own ideas and viewpoints. My best students have all ended up teaching me a lot more than I ever taught them. This process only comes about, however, if we allow them room to breathe and make their own mistakes. 
\item {\em Scientific competition – or working in a crowd.} One result of the intense competition that I encountered during my Ph.D. work was that I swore that, never again, would I  consciously enter a ‘popular’ field in which numerous other researchers were working. If a lot of good people are working on a problem, my own contribution will generally be superfluous. I can make a lot more progress (with a lot less stress) if I can choose a question that no one else is looking at. This will also give me a chance to sit down and think, without the paranoia that someone else will ‘scoop’ me if I do. It certainly helps that the question that one chooses to look at is a good one, but often one won’t know that until he/she has started.
\item {\em Good research is when you are surprised.} I know that my own imagination is much more limited than nature’s. The most important thing that a scientist should remember is to keep his/her eyes open; nature will provide often unexpected answers, if we are ready to listen. Expect (hope) to be surprised. In addition, never believe yourself too much. Keep in mind that we are all idiots, so always double, triple and quadruple check yourself. Sanity checks are critical, and are necessary to ensure that you are really listening to nature – and not to your own wishful thinking.
\item {\em Different modes of doing science.} In my Ph.D. work, I was exposed to 3 different types of experimental science/scientists. These are: (1) those that will only go forward when a theory, with definite and well-defined predictions, exists (2) those who dig ever so more deeply on a single problem/question with the intention of achieving comprehensive understanding (3) those who blindly dive into to a question where no framework for understanding it yet exists. All of these types of scientists are necessary for science to advance. Personally, if you have not yet guessed, my own preference is for the science embodied in (3). Scientists (especially theorists) are infinitely more limited in what they can imagine than is nature. If our job is to try to understand how the world around us is composed, we should not proceed wearing blinders. For this reason, we shouldn’t be afraid to jump into a scientific maze, even if we don’t have any idea ahead of time where it will lead us. The questions and science will come, so long as we keep our collective heads up and our eyes open. Often, those that choose path (2) may also become like we 3ers. I have seen cases where, by digging ever more deeply with ever more precision, people have unexpectedly uncovered new worlds. The unexpected is waiting for us, so long as we are positioned to see it. 
\end{itemize}

\section {Experiences as a postdoc}
After completing my doctorate, my wife, Tamar, and I had to decide how to proceed in life with our small,  but growing family.  We decided that, if the opportunity presented itself, we would try our luck in post-doc land. We made this decision for all of the right reasons. I never anticipated that I would continue on as a researcher. I found that I liked banging my head against the wall, and trying to put together insurmountable problems. I never, however, really thought that this would lead to an academic career. Basically, the rationale in continuing to postdoctoral research was as follows: (1) I liked to play with physics (2) we were, as a family, young and curious of how life takes place in other places. We, therefore, decided to give it a try. I would continue on with my head-banging for a bit longer, and, together, we would experience something new. Life is a long journey through a maze that has only one end to it, so it made sense to us to take a break from ‘real life’ for a little while longer. As it turned out, I never really had to work for a living, but at that time we had no inclination (or expectation) that this would turn out to be true.
	
 We ended up at the University of Texas, quite by accident. I had been corresponding for some time with a well-known researcher about the possibility of joining his group. I had more or less assumed that we would go there until, surprisingly (for me), I received a message that he had decided to take someone else.  
 
 This was quite late in the game. With nothing to lose, I sent out a flurry of CV’s etc (by snail mail) and, generally, received a pile of standard rejections stating, politely, “don’t call us, we’ll call you”.  I, eventually, received two offers. I was in the process of emailing a positive reply to one of these (in fact I was sitting in front of my screen about to press ‘send’) when the phone rang. Picking it up, I heard the hissing of a long-distance call and, from the other side, came out something totally unintelligible. The connection seemed to be fine, and the caller seemed to be offering me a job in a language that was vaguely reminiscent of English. The caller’s accent was so thick that I couldn’t be sure. After  a few times of “can you repeat that”, it turned out that the caller was none other than Harry Swinney, from the University of Texas at Austin. Harry was offering me, in his very thick Louisianan accent, a postdoctoral position at a salary slightly larger than one that I was about to accept. Never having been to Texas, I was a bit reticent about riding a horse to work, but I figured that the extra income might come in handy, as our two toddlers were showing signs of becoming increasingly hungry. As a result, for all of the wrong reasons, I made the best decision that I could possibly have made, by telling Harry yes. {\em Moral: luck is crucial in science – immeasurably more important than common sense}.
	
I won’t mention much about our personal experiences in Texas except to say that we stayed there for nearly 4 years and had a tremendous experience. I will say that, by an informed source, I was informed (prior to our arrival) that Harry was Jewish. Coming from Israel, this made me feel that, in some sense, we had some family in this unknown place. Moreover, the mere fact that Harry was Jewish had a major impact on our personal relations. Upon arrival, everyone would refer to Harry as “Dr. Swinney”, whereas I immediately called him by his given name, “Harry”, since we were from the same tribe and all. Harry, seemed to have no problem with our informal relationship and, in fact, invited our little family to his house for Thanksgiving dinner.  When, during grace before dinner, there were a few references to “our Lord Jesus” Tamar and I exchanged surprised glances (“what kind of a Jew is this?”). It turned out that our `informed' source had been wrong (when later asked about it, I received the reply that “someone that small HAS to be Jewish”. I note that Harry’s scientific stature is quite a lot larger than his physical one, but my source is even shorter!). On the other hand,  our easy and familiar relationship was set and continues in the same vein to this day; we count Harry and his family as close friends. 
	
What about science in Texas? The scientific atmosphere there was fantastic. The Center for Nonlinear Dynamics was composed of a number of faculty, 3-4 postdocs and a lot of graduate students. Harry had established the center a few years before and it was considered one of the world’s Meccas for nonlinear science. There was an exciting and very friendly atmosphere that pervaded the group. (We all came to work by bikes or cars – no horses, to our surprise). This unique and casual atmosphere essentially created an extended family for us. 
	
Harry’s style of management was diametrically different than what I was accustomed to in Victor’s group. Whereas Victor was very hands on, Harry was very much ‘hands off’. Essentially, he recruited the best people that he could find, made sure that they had what they needed, and let them work. We would have weekly or biweekly meetings to keep him up to date, but, as postdocs, we were provided a unique opportunity to “sink or swim”.  When I first arrived, Harry suggested one or two projects that I might be interested in pursuing. After a week or so of wandering around the lab looking into them, I decided that none  really interested me. To his great credit, Harry didn’t get upset with this “chutzpa” (both of us, at the time, being Jewish). He gave me free reign to continue to wander around the lab to see if I could come up with something that I wanted to do. 
	
	While wandering around and talking with people in the group, I had the great fortune to meet Mike Marder. Mike was, at that time, a new faculty member, who had just come to Texas from his stint as a postdoc at the University of Chicago. While there, he had heard a talk by an expert in the fracture of ceramics who had described a rich phenomenology of effects (note: there is that word “rich” again) that commonly are observed when things break. Moreover, Mike, upon arriving at Austin, had rented a “rent a wreck” car that had a large developing crack on its windshield. I guess that he was doing a lot of driving around at the time, with this crack in front of his nose. This started him thinking, as the crack seemed to be forming a cusp-like structure. Mike, immediately, came up with the theory that slow cracks were spontaneously able to form singularities, such as cusps. Mike was convinced that these ideas provided fertile ground for an experimental foray into the dynamics of slow cracks. Not only was Mike convinced about this, he made it his mission to convince me as well. 
	
	Mike was incredibly persistent. So much so, that I felt that to shut him up, I would have to perform some kind of experiment. I was, in fact, pretty convinced that he was wrong about the cusps (as I could see no reason, a-priori, for their formation).  On the other hand, Mike is one of the smartest people that I have met, and I figured that, since I had nothing better to do, I might as well put something together to check whether there was anything worth seeing in this business. 
	
	Without knowing anything about fracture or cracks, I went down to the machine shop and constructed the following `high-tech' apparatus. It consisted of plexiglass (PMMA) plates, each with an initial `seed' crack sawed at the center of one of the sheet’s vertical edges. At the top and bottom of each plate, I glued plexiglass tabs. I affixed the top tab to the lab’s ceiling via a steel cable. To the bottom tab I affixed a steel basket, which I loaded up with  numerous lead bricks that I had found scattered around the lab. This was my loading apparatus. Now I needed a measurement device. I borrowed an old camera and videotape recorder. I focused the camera on the vicinity ahead of the seed crack. I now needed a trigger for this state-of-the-art experiment. The idea was to look at slow cracks moving across the material and to convince Mike that nothing interesting was going to happen (e.g. spontaneous formation of a cusp).  I had a lot of time on my hands, as I was trying to figure out a `real' project to work on. I therefore decided to let gravity do its work. I turned on the video recorder (it could record for up to 4 hours) and suspended the whole assembly at a height of about 1/2 meter above a metal drum that was sitting around in Harry’s lab. I then went to putter around the lab, while trying to come up with interesting ideas for experiments. Whenever the sample would break (as would eventually happen), the sound of 100kg of lead bricks falling on top of a steel drum would reverberate throughout the lab. If, by chance, I didn’t hear it, someone else in the lab was sure to let me know. (I think that people started taking bets on if and when a sample would break.) I would then rush to the scene and turn off the video tape recorder. I would then go through the recording, frame by frame, to try to follow the crack’s dynamics. 
	
	Invariably, I would find that in one frame the crack was there, and in the successive frame (1/60 of a second later) the material had broken, with the top half of the sample dangling in front of the camera. As I am a stubborn guy (actually Mike was even more stubborn and refused to stop bugging me), I set about varying the number of bricks, changing the length of the initial crack, and trying to perform various voodoo dances around the steel drum, both before and during each experiment. Despite my many efforts, exactly the same scenario would repeat itself; one frame the crack was there whereas in the next frame the whole sample was broken.
	
	I must have repeated different versions of this scenario 30 or 40 times. I was never able to visualize a single fracture event. I did, however, save and label the broken pieces of plastic, since I’m a firm believer in never throwing anything away (this characteristic trait is not, necessarily, viewed as an advantage at home). In the lab, however, innate stubbornness when coupled with the hoarding of junk sometimes pays off. I noticed that on each of the broken surfaces formed by the crack that created them, there was an ordered pattern. Its existence seemed to be independent of the applied load, while its (millimeter scale) wavelength seemed to only slightly vary with the number of lead bricks (all else being equal). Furthermore, the fracture surface immediately after the seed crack seemed to be mirror-like, with no pattern at all. The pattern would only show up a few centimeters beyond the seed crack. 
	
I remind you that I had just finished my Ph.D. work in investigating pattern forming systems (albeit in cognac). I also remind you that, at this stage in Harry’s lab, I had nothing better to do. We then immediately set about trying to figure out where this particular pattern was coming from. As neither Mike nor I had much knowledge about fracture at the time, we set up meetings with a number of faculty members in both the neighboring engineering department as well as some big-name people in neighboring institutions. We discovered that people had seen this stuff on and off, but no one who we talked to seemed to think it was particularly important or interesting.

 The fact that people had observed this stuff in other materials spurred us on. In addition, the overall message that we were receiving seemed to be that the only thing that the field of fracture needed was better numerical codes. Understanding a bunch of insignificant markings was neither interesting nor a priority. To a physicist, hearing such comments was like waving the proverbial red flag in front a bull. We immediately felt that there was something fundamental that needed to be understood. Mike and I also realized that, apparently, we needed to figure out a better way to look at this stuff than videotaping it. We started reading the literature (always a good idea, unless you are deeply eager to rediscover the wheel) and discovered, among other things, that cracks could propagate to nearly material shear wave speeds. This explained why the videotape hadn't provided much insight. We also understood that, if we wanted to understand these patterns, we apparently needed to study {\em fast} (or ‘dynamic’) fracture. Doing the math, we realized that to measure a crack’s instantaneous speed at the spatial scales capable of resolving our millimeter-scale patterns, we had to come up with a way to measure the crack’s location continuously, at up to 10,000,000 times a second (a bit faster than the 60 times a second that the videotape could provide). 
 
 I didn’t have any experience working at these data rates (remember my convection experiments took place over weeks and months), but, as the saying goes, “fools jump in where angels fear to tread”, so I, of course, jumped in. With the help of my friend (and, at the time, graduate student) Steve Gross, we set about putting together a measurement system (on a shoe-string budget) that was both faster and orders of magnitude more precise than anything built before. All of this was needed to resolve these insignificant markings on fracture surfaces. We accomplished this by harnessing a good idea with (you may have guessed it) new technology. 
 
 Rapid digitizers that interfaced to personal computers were just coming out. We purchased one of the first of these, and immediately set to work writing (nonexistent) drivers for it. We monitored the crack progression by coating the materials to be fractured with a sub-micron electrically conductive layer. When a crack progressed, it would cut this layer and thereby alter its electrical resistance. This resistance was measured by the fast digitizers (after amplification by home-made electronics) and, after calibration, supplied us with the crack’s instantaneous location every 0.1$\mu sec$. 
 
 I won’t go into any more detail, but our attempts at trying to understand these ‘insignificant’ surface markings have led to a new understanding of crack dynamics (or how brittle materials break) \cite{Marder_howthingsbreak_1996}. Incidentally, new forms of numerics actually had to be developed in order to follow the resulting dynamical behavior that our experiments revealed. An interesting aside is that our first paper, which presented a description of an instability of cracks that causes the patterns that we observed \cite{Fineberg_instabilityPRL_1991}, was nearly impossible to publish. Only one of the referees consulted thought that the work was, at all, interesting. An additional referee finally agreed to recommend publication while stating something like “I don’t think that this is really that important, but it’s a good idea to get physicists involved in studying materials, so I’ll recommend it”. As this paper has since been referenced hundreds of times, someone must now feel that the work was worth something. 

At this juncture, I wish to make some general comments:
\begin{itemize}
\item	If you believe in something, go for it. Trust your ‘gut feeling’. Even if your intuition turns out to be a bit off, you may still find interesting new physics. This field was nearly nonexistent in Physics when we started working in it – and this is a good thing.  The lack of competition enabled us to work slowly and carefully, while only publishing when we had something to say.
\item	Technology, again, was critical here. Even had someone wanted to study the fracture patterns, it would have been impossible to achieve without the technology that then had become available. For this reason we were fairly certain that we were not entirely ‘re-inventing the wheel’, while breaking into the very mature field of fracture mechanics.
\item	Don’t be discouraged if you have a really new idea and can’t get it either published or funded (it took us 5 years or so to get funding for this research). Truly new ideas will generally generate resistance if they go against common wisdom. If you do your homework, and are sure that you are right, then a good idea is worth fighting for. No new idea that I ever came up with (I’ll mention a few more later) was ever funded at first. Generally, I had to divert funds from other (more mainstream) research directions to fund really new research. It often only takes a single negative referee to kill a proposal and a truly new idea will generally be doubted by people who may be entrenched in ‘accepted ideas’.  If you, however, continue to bang your head against the walls defined by pre-conception, they will often, in the end, budge. 
\end{itemize}

While at the University of Texas, I was also involved with another major project, with both Harry and then graduate student, Dan Lathrop.  We constructed what became known as the `Texas Washing Machine' to study the transition to high Reynolds number turbulence. This was, philosophically, a different type of experiment than the fracture experiment. Instead of having our curiosity piqued by some unexplained phenomenon, here our intent was to drive a well-known nonlinear system well beyond regimes where its behavior was known. This was a wholly exploratory experiment, where we had no real idea of what we would find. Dan was the real driving force for this experiment, and its construction was an engineering coup. 
	
 Without going into too many details, the general idea was to probe the unknown regime of very high Reynolds numbers in a very large Taylor-Couette apparatus.  Taylor-Couette flow is the flow of  fluid confined between two co-axial cylinders. In our case, the inner cylinder was rapidly rotated while the outer one was stationary.  This is another experiment  for which we did not have a clear theoretical idea of what we would see. A lot was known about its nonlinear (pattern-forming) behavior at low to moderate Reynolds numbers, but little beyond. We decided to measure a new type of order parameter; the applied torque needed to achieve a given Reynolds number characterizing the flow. We realized that this was the type of experiment that breaks new ground; driving a paradigmatic flow to unprecedented levels. While, a-priori, we had no idea what we would find, it was not entirely surprising that this experiment uncovered interesting new results; an unexpected transition at high Reynolds numbers to a turbulent flow dominated by the shear layer adjacent to the cylinder walls \cite{Lathrop_Fineberg_1992}. Having no real preconceptions of what we would see, we simply kept our eyes open – and enabled the new physics to reveal themselves. 
  
\section{Life as an independent researcher}
Towards the end of our stay in Texas, it turned out that there were a number of academic positions that were being offered to me. I really didn’t expect this to happen, but when it did, and a job at the Hebrew University in Jerusalem was being offered, we didn’t hesitate. Both Tamar and myself are from Jerusalem, and the thought of having two sets of grandparents who actually wanted our (now 3) kids was irresistible.

 I had three different projects in mind when I sent in my research proposal. The first project, was to continue my work in fracture. A second project was to follow the detailed dynamics of crystalline defects to examine how crystalline order gives way to disorder as the effective temperature is increased. The third project was a fluid experiment that I never really got around to starting.
 
To perform the second project I found, from the literature, that it was possible to create crystalline patterns on a fluid surface by exciting waves via a parametric resonance. To put it simply, by shaking a fluid layer up and down, one could create waves on a fluid’s surface. By mixing the shaking frequencies and amplitudes, crystals of very different lattice structures could form. I intended to generate disorder in the system by controllably increasing the shaking amplitude (or “temperature” of the system). I got the idea of forming the crystals from beautiful experiments performed by Stephan Fauve’s group, located at the time at ENS Lyon.  A friend of mine, Stuart Edwards (formerly in the Texas group), had joined Stephan’s group as a post-doc. Stuart had shown that even quasi-crystalline patterns were possible \cite{Edwards_1994}. 

Stuart and Stephan very generously invited me to visit for a week to Lyon, to teach me the tricks of the trade (as I had no prior experience in this type of experiment). Stuart taught me that, to ensure that the system’s boundaries did not enforce their symmetry on the crystals, it was necessary to increase the dissipation in the fluid layer. Stuart and Stephan had cleverly demonstrated this by generating quasi-crystalline patterns on the surface of a container whose boundaries were shaped like the outline of France (personally, I would have created something simpler, but I have never been the artistic type).  The fluid dissipation could be increased by either increasing the fluid viscosity or reducing the height of the fluid layer. As I was interested in the intrinsic disorder mechanisms of the  crystals, I decided to both increase the fluid viscosity {\em and} reduce the fluid layer height to a minimum. 

It took over a year to set up the surface wave system, with the homemade software and electronics needed to drive it, monitor it, and visualize the fluid surface in real time. When we were about to crank it up for the first time, I had to leave my lab for a week to attend a conference. While I was away, I asked my graduate student, Oleg Lioubashevsky, to ‘go wild’ and explore the phase space to see types of nonlinear crystalline patterns we find.

Upon my return, I asked Oleg what he discovered. His answer was something like ‘nothing much’.  When Oleg told me that he barely saw any patterns in the system, I asked him if he tried to increase the forcing. His response was that he had driven the system to about its limits. I suspected that our imaging system probably needed adjusting, since we should expect to see {\em something}.
 
Oleg had adjusted the camera to be focused on the center of the system, at the expense of its lateral edges, since we were only interested in pattern dynamics that were unaffected by the system boundaries. Moreover, to prevent effects of stray light on the image quality, Oleg had conscientiously put up baffles that blocked direct optical access to the fluid surface. I was a bit confused, and had begun doubting the optical set up that we had designed. To get a better look, I removed these baffles and took a look “under the hood” of the system.
  
We immediately encountered one of the “wow” moments, that sometimes make all of the frustrations of doing experiments seem worthwhile. Around the circumference of the (circular) lateral boundaries of the system, we saw huge amplitude, highly localized waves running around and around the system’s lateral boundaries. These were some of the strangest nonlinear creatures that I had ever observed. In fact, we started to call them our ‘Loch Ness Monsters’ as they indeed looked like plunging sea snakes, diving in and out of the fluid. These highly localized propagating waves, that we officially called ‘dissipative solitons’, became the basis of Oleg’s thesis work \cite{Lioubashevski_DissSolitions_1996}. In fact, their mutual and self-interactions by means of the wake that they created, possibly inspired Yves’ later work on ‘walkers’, particle-wave analogs whose properties were reminiscent of the pilot waves envisioned in early quantum mechanics \cite{Couder4}.

	The Loch Ness monsters’ close cousins,  localized non-propagating states, were discovered slightly afterwards by Paul Umbanhowar, then in Harry’s group, within vibrated granular systems. This new family of huge amplitude stationary waves were coined ‘oscillons’ by Harry’s group \cite{Umbanhowar_oscillons}. Oscillons, together with our own Loch Ness monsters served to ignite the imagination of the nonlinear community. My group’s later work (performed both by Oleg and my student, Hagai Arbell) revealed the close connections between these weird types of waves \cite{Arbell_Oscillons_2000,Lioubashevski_Oscillons_1999}. We did, however, make a bad public relations mistake in giving the Loch Ness monsters such a boring name (‘dissipative solitons’). Paul’s term, ‘oscillons’ was certainly a much catchier moniker. The splash that they created (pardon the pun) was initially much greater, as the catchy name  caught on like a match thrown into in a dry forest. (Some unsolicited advice: if you discover something really new, it is a good idea to invest some time in naming it.)
	 
	Beyond the study of Loch Ness monsters, this surface waves system provided numerous insights on how, why and where nonlinear waves interact. This  work was initiated by Hagai Arbell and continued by Tamir Epstein. By driving the system with two or more frequencies simultaneously, we found that nonlinear waves of different wavelengths interact strongly with one another, and form fascinating patterns of high amplitude waves. The dynamics of these waves, in both space and time, fired the imagination of numerous experimental and theoretical groups. The beauty of this work was (1) it really was aesthetically beautiful and (2) it was a relatively simple and extremely well-controlled realization of a very general and universal question: `What are the ways by which nonlinear waves can interact?'. We had no idea of the answers that nature would supply to these questions, but, from the start, we understood that this was a question that was certainly worth posing. Our experimental system was the simplest that we could conceive of that could enable nature to provide us with a clear answer. I'm  quite proud of how this turned out \cite{Arbell_PRE_2002,Epstein_PRL_2008}. 
	
	The discerning reader may have noticed that the aforementioned projects actually had  very little to do with the original reason that I started this research direction, which was understanding the dynamics of how crystalline materials become disordered. Three very successful Ph.D. theses were concluded solely by studying interesting and wholly unexpected phenomena that we encountered along the way. In my view, this is the way that pure science should be pursued. Keep your eyes open and let nature guide you. Don’t make the mistake of trying to lock onto a narrow goal at the expense of more interesting phenomena that you may encounter. Eventually, I actually did end up exploring how 2D crystals break up when sufficiently pushed. This work, performed by my then M.Sc. student, Itamar Shani, was certainly interesting \cite{Shani_PRL_2010}, but I don’t believe that its results had the impact of the many detours that we encountered. Curiosity-driven research often reveals the surprises that, for me, make the journey worth taking. 
\subsection{Breaking stuff for a living}
	In parallel to our work on surface waves, my group continued the work on fracture that I had started in Texas. This manuscript, by the way, is getting a bit lengthy. I had no idea that my ramblings would go on for so long.  My first student, Eran Sharon, performed beautiful work in both characterizing the instabilities that a crack undergoes and connecting these new phenomena to the existing theories of fracture embodied in the field of fracture mechanics. Eran demonstrated that the ‘insignificant’ surface patterns that first piqued our curiosity in crack dynamics actually have a huge effect on how and why a crack will dissipate energy; the fracture surface patterns are, literally, the tip of an iceberg. We found that the fracture surface patterns form a complex network of sub-surface microscopic cracks \cite{Sharon_PRL_1995} beneath them (we coined them ‘micro-branches’ – not quite as good as oscillons – but certainly catchier than ‘dissipative solitons’).  Once excited, micro-branch networks increase the energy dissipated by a moving crack by over an order of magnitude as they accelerate \cite{Sharon_energy_1996}. 
\subsubsection{	Crack front waves and meteor impacts}
	Eran also noticed that grooves of infinitesimal (10-100nm) amplitudes seemed to form spontaneously on mirror-smooth fracture surfaces, before micro-branches initiate. It was entirely unclear what their origin was, but when looking closely enough, they always seemed to be around. Furthermore, they were long-lived and were always observed at well-defined angles to the direction of crack propagation. Intrigued, Eran devised a way to measure these waves and relate their appearance to the overall crack dynamics. Entirely by luck, Eran attended a conference on a beach in France, where Jim Rice gave a talk about a new type of wave, ‘front waves’ that, at least theoretically, appeared to live on crack fronts \cite{Morrissey_Rice_2000,Ramanathan_Fisher_1997}. Front waves are formed by the interaction of material defects with crack fronts, as they momentarily distort the singular leading edge of the propagating cracks. Waves excited by these interactions were predicted to live within the singular crack front and persist over long distances. Jim’s theory struck a chord with Eran. Upon returning from France, he immediately initiated experiments that quantitatively verified the predictions of this new theory. In addition, these experiments suggested that front waves actually seemed to behave as a type of nonlinear solitary wave; highly localized in space with very little dispersion as they propagate \cite{Sharon_FWNature_2001}.
	
	The work on front waves provided the key to understanding the geological phenomenon of ‘shatter cones’. Shatter cones are rocks having a, highly curved, shape whose surfaces are decorated by dense patterns of V-like striations, all having the same angle. Shatter cones, which are independent of the rock type on which they are formed, are only observed near locations where large meteors have impacted the earth’s surface.  Our work, with Amir Sagy and Ze’ev Reches \cite{Sagy_Nature_2002,Sagy_ShatterJGR_2004},  revealed that shatter cones are actually huge micro-branches formed by the enormous energy imparted by meteor impacts. The characteristic striations were no other than front waves that formed along these crack fronts, whose propagation velocities approach the highest allowed crack velocities.  The story of nearly invisible surface markings $\rightarrow$ front waves $\rightarrow$ shatter cones is another nice example of how understanding a seemingly innocuous observation, that for whatever reason piques one’s curiosity, has lead to discoveries whose ramifications extend well beyond anything that could have originally been anticipated.
	
\subsection{How an unfunded biophysics experiment led to new understanding of fracture, friction, and earthquakes}
	A few years after arriving at the Hebrew University, a good friend of mine, Alex Levine, was recruited to our biology department. Alex and I had been together in both high school and in our obligatory army service and it was great to catch up with one another. As our coffee meetings at the University faculty club started to be increasingly frequent, we decided that we ought to try to justify them  scientifically. Since Alex was an expert in molecular biology and I was busy breaking things, we decided to set up an experiment designed to break single biological molecules. We decided to start with DNA and work ourselves up to mechanically unfolding proteins to better understand their structure and dynamics. After a sufficient amount of consumed caffeine  (over a period of a few months), we came up with a whole research plan that even included attempting to study correlations between the efficacy of certain anti-cancer drugs to the strength of their mechanical attachment to DNA and other bio-molecules. To this end, Gil Cohen (my then and current research scientist) and I put together a really nice experimental system in which we tethered single DNA molecules between a glass plate and magnetic beads. By pulling on the beads with strong magnetic field gradients, we could apply a controlled range of forces to unravel the DNA until actually breaking it. In parallel, we implemented an evanescent imaging technique in order to measure the separation of the bead from the glass to nanometer resolution. 
	
	Unfortunately, at the time, we couldn’t get any funding to carry out these experiments. I still have the series of failed research proposals, including the negative reviews of numerous learned referees, that told us in exquisite detail why these measurements were both impossible (while we were actually doing them) and, furthermore, uninteresting. In the end, I gave the project up, as I didn’t have the means for supporting the research. To this day, 25 years later, I believe that these ideas had (and still have) significant potential. Many of them have, since, been successfully performed by other people. Some of them, so far as I know, still haven’t been tried.
	
	While this project was never completed, it actually contributed, in a major way, to some entirely new and rather fruitful research directions. These are (1) fooling nature to enable the study of fracture with unprecedented detail; (2) establishing a new and fundamental paradigm for friction and (3) obtaining a fundamental understanding of earthquake dynamics. I suppose that this study goes under the category of “when life deals you lemons, make lemonade”. 
	
\subsubsection{	Understanding fracture dynamics by breaking “scientific jello”}
	During our days of pulling apart DNA, we needed to obtain well-defined lengths of certain DNA fragments. To achieve this, we used the standard biological technique of electrophoresis; pulling DNA fragments with an electric field through an obstacle course constructed using aqueous polymer gels, made of polyacrylamide. During this process, some of these rather slippery gels were accidentally dropped. We noticed that, when this happened, the gels could typically shatter very much like a plate of glass falling on the floor. These gels are very compliant materials. As their elastic moduli are, typically, about a million times more compliant than glass, their elastic waves are about 1000 times slower.  Crack velocities scale with approximately the shear wave speeds of the medium. We surmised that crack dynamics within gels could, therefore, be slowed down by a factor of 1000, from kilometers/sec to  meters/sec. We realized that combining this with fast photography, it would be possible to follow crack dynamics with unprecedented resolution and detail.  At about the same time, a friend, John Bechhoefer, visited us and suggested the same idea. John even showed us papers describing how to tune the gel properties by playing with their chemical composition.  We then decided to give it a try. 
	
	This lead to the Ph.D. thesis of Ariel Livne, who started playing with this scientific ‘jello’ as an undergraduate project. Ariel first showed that all of the behavior that we observed in ‘standard’ materials takes place, in exactly the same way, within brittle gels \cite{Livne_PRL_2005}. Moreover, by means of fast photography we could follow the shape of the singular crack tip, for the first time, {\em during} its propagation.  Comparing these crack tip shapes to those predicted by fracture mechanics, we immediately found that, very near the crack tips (precisely in the region that we expected fracture mechanics to work best), these predictions seemed to fail.  
	
	These results were so surprising, that we spent a good deal of time double-checking them, to make sure that we didn’t make any mistakes. By constraining our analysis to scales a few mm’s away from the crack tip, we found that the predictions of fracture mechanics seemed to conform well to quantities which could be independently measured, such as the fracture energy (energy needed to create new surfaces).  We found that we could measure the strain fields surrounding the crack tips by looking at deformations of densely imprinted scratch patterns on gel surfaces, as cracks propagated. There too, we found agreement with fracture mechanics away from crack tips, but increasingly large discrepancies the nearer that we got to their tips. 
	
	About the same time that we were getting increasingly confused, Eran Bouchbinder joined our group as a postdoc. Eran is a theorist, and I really had no idea what to do with him.  Before he joined us, I had strongly tried to convince him to go elsewhere, to do a postdoc where he could learn something. Eran, however, told me that he was willing to take his chances. He wanted to go to a place where the data was being generated. He didn’t want to hear about experiments ‘second-hand', when results had been either consciously or subconsciously filtered. Well, we were certainly generating data. 
	
	Once we all convinced ourselves that our data were really valid, Eran set about coming up with ways to explain what we were seeing. This ended up in the first fundamental extension of fracture mechanics in decades. Eran, together with Ariel, showed that all of our measurements were in perfect quantitative agreement with a new theoretical description, derived by Eran, that incorporated nonlinear elasticity near the crack tip \cite{Livne_PRL_2008,Bouchbinder_PRL_2008}. In hindsight, this theory made perfect intuitive sense. (Linear Elastic) Fracture mechanics was based on the assumption that materials are linearly elastic (i.e. obeyed Hooke’s law). This assumption, while perfectly valid for small strains, has to break down when strains are large enough to pull apart the material. This, of course, is precisely what is taking place at the tip of a crack! In her PhD work, Tamar Goldman extended these experiments in gels to demonstrate, for the first time, the validity of the equation(s) of motion for fast cracks \cite{Goldman_Interia_PRL_2010} (one derived by Freund for infinitely large systems and another derived by Mike Marder for cracks propagating within a finite strip \cite{Marder_1991}. At the end of the day, breaking ‘jello’ enabled us to obtain an excellent theoretical understanding of the form and dynamics of `simple' cracks \cite{Livne_Science_2010,Bouchbinder_AnnRev_2010}. 
	
	What about non-simple cracks? Tamar and Eran utilized the newly developed nonlinear theory of fracture to quantitatively explain spontaneous crack oscillations that develop at very high fracture velocities, attainable when micro-branching is suppressed \cite{Goldman_Oscillations_2012,Bouchbinder_Goldman_2014}. These observations in brittle gels have gone on to become a testing ground for new ways to compute rapid crack dynamics, as recent work coupling phase field descriptions to nonlinearity near the crack tip has shown\cite{Karma_bouchbinder_NatPhys_2017}. Finally, Tamar’s work in gels, using microscopic grids imprinted on gel surfaces to directly measure strain fields, has also provided a fundamental description for the micro-branching instability \cite{Goldman_microbranchin_PRL_2015}, that had stubbornly evaded explanation since we first observed it in Texas (remember how we got into this business in the first place). More recent work by Itamar Kolvin, utilizing thick sheets of gels, has directly observed crack front dynamics in real time. This has provided us with both a quantitative description of front dynamics \cite{Kolvin_PRL_2015,Kolvin_Mokhtar_PRL_2017} and the first description of how facets (that look like cleaved planes) spontaneously form on the fracture surfaces of {\em amorphous} materials \cite{Kolvin_Nat_mat_2018}. 
	
	Polyacrylamide gels can also be made to be extremely tough, by introducing a second, loosely cross-linked chemical network alongside the first \cite{Gong_2003}. These gels stay intact for strains an order of magnitude higher than those that would fracture unenhanced polyacrylamide gels. Using similar experimental methods, Itamar and John Kolinski demonstrated how the character of fracture changes in these super-tough materials \cite{Kolvin_Kolinsky_PRL_2018}. This is another example of when `fooling' nature down can provide insights into questions that couldn't be considered otherwise.  
	
	\subsection{ The fundamental physics of friction and earthquake dynamics}
	Since I had heard of the concept of friction, I never really understood it. The reason is as follows. Let us consider the contacting surface that connects any two sliding bodies. The laws of friction tell us that the onset of sliding of these surfaces can be described by a single number, the static friction coefficient. This concept is so entrenched that we generally accept it as dogma. As established previously, I have always been a contrarian. If something is generally accepted as truth, I often will unconsciously start to question it. The theory of friction has always bothered me, in this sense. It always seemed too good to be true. By the following thought experiment, at some level, I believed that this idea had to be wrong. If we consider a spatially extended frictional interface, I never understood how all of the frictional contacts that compose the interface would decide to detach at the same time.  This, however, is precisely the process that is embodied in the idea of a `friction coefficient'. 
	
	This idea was sufficiently confusing to me, that I decided that it was worth putting it to a test. To do so, we would have to figure out a way to measure when contacting points actually detached from one another. The answer turned out to be an aftermath of our ‘frustrated’ experiment in biophysics. There, we utilized evanescent imaging to track the extension of biomolecules; if we illuminate a surface by a sheet of light whose incident angle is well beyond the angle for total internal reflection, the amplitude of scattered (evanescent) light above the surface decays exponentially with the distance of the scatterer from the surface. After about 50nm of separation, the light’s intensity decays significantly. Gil Cohen suggested that we could utilize this to quantitatively measure the real area of contact at each point along a frictional interface. Furthermore, we realized that with a sufficiently fast camera, this could be done in real time.  
	
	 My M.Sc. student, Meni Shai, designed a home-made rig that could controllably slide long plates of plexiglass over one another. My then student, Shmuel Rubinstein, started to implement our imaging ideas. Before committing ourselves to complex and expensive contact measurements, we decided to perform a `proof of principle' measurement by simply using two laser beams to evanescently illuminate the interface at locations separated by about 100mm. The output of the transmitted laser beams was continuously detected by two photodiodes that were interfaced to a 2 channel digital oscilloscope. Any change in the intensity of one of the beams would stop the acquisition, enabling us to see whether changes in the beam intensities took place at the same time or not. We were really shocked (and a bit excited) to see that the photodiodes reproducibly `blipped' upon the onset of frictional sliding, with the two blips repeatedly separated in time by about 100$\mu sec$ (a 'wow' moment). The system was really telling us that contact separation was not simultaneous, as we suspected. In addition, the time difference corresponded to about the speed of about 1000m/s, about the speed of a shear wave (or crack).
	 
	   This preliminary result pushed us to develop the full imaging system. We couldn’t find a sufficiently fast camera at the time, but we saw that a sensor that could do the job was just coming out. The electronics for our first fast camera were actually home-made and housed in a shoe box. Luckily, soon after, a commercial camera using the same sensor became available.  Our first results \cite{Rubinstein_Nature_2004} clearly showed 3 different families of `detachment fronts' that broke the frictional contacts to enable the start of sliding. We were really excited by these results. Eventually we understood that we were really looking at earthquake dynamics (the frictional slip of tectonic plates). Surprisingly, instead of reinventing the wheel, our experiments were providing very new insights. Seismic data, the bread and butter of the earthquake business, could not provide the type of detail about earthquake dynamics that we could directly observe in the laboratory. 
	
	We have since performed a lot of work in this direction. Below are some high (and low) points in these studies. 
	
	 {\em Sometimes frustration pays off}. 
	 
	 My student, Oded Ben-David, succeeded Shmuel in the friction experiments. Shmuel had originally intended to finish his Ph.D. work quickly. As Shmuel’s progress was really rapid, I had recruited Oded to take over within a few months of starting his Ph.D. project. Shmuel, however, for personal reasons decided to extend his doctorate for an additional year or more. As Oded liked to build things, we decided to construct a new (and considerably better) experimental apparatus in order to take the next step. Oded set to work and, in record time, constructed a much improved apparatus; it was a better controlled and considerably more rigid version than the previous one. Oded then proceeded to take data. To our surprise (and consternation) Oded found that he couldn’t easily reproduce Shmuel’s experiments, despite the fact that Shmuel could do so at will. The harder that Oded worked to perfect his new apparatus the more random his results appeared. The random nature of Oded’s experiments expressed itself in both the locations where earthquakes would nucleate and their speed or `family'. Even the apparent ‘friction coefficient’ in the experiments would randomly vary by factors of two or more.
	 
	  The harder Oded worked, the more frustrated he became. Finally, after nearly 3 years with inconsequential results, we realized that we were apparently missing something very basic. This was a classic example of futilely beating our heads against an immobile wall. Something had to give, but it was, apparently, not going to be the wall. Out of pure frustration, Oded decided to tile the entire frictional interface with strain gages. This was basically a sanity check to verify that the stresses that we were remotely applying and assuming to be spatially homogeneous were indeed so. Oded was both really frustrated and, by this time, a genius in electronics, programming and mechanics. Within a week he had the new system running (100 strain gages mounted, measured in parallel and analyzed), although I don’t think that he slept much that week. 
	  
	  What Oded discovered was truly eye-opening (another `wow' moment). While the applied stresses were as uniform as we could make them, both the normal and shear stresses at the interface were anything but uniform! Oded discovered that even the slightest non-uniformity in system alignment (or even sub-micron ‘bumps’ in the contacting surfaces) had a profound effect on both the spatial stress distributions and, importantly, on the type of earthquake that we would excite. Armed with this new information, he then went on to make sense out of all of the seemingly disparate results that he had obtained, explaining why, how, and where each type of earthquake along our laboratory ‘fault’ took place \cite{Bendavid_Science_2010}. This work produced powerful insights for the dynamics of earthquakes along natural faults as well, highlighting the profound effects that even slight inhomogeneities produced in a ‘near perfect’ laboratory system. 
	  
	   Among other fundamentally important results obtained in these experiments, Oded went on to demonstrate that the idea of a characteristic ‘static friction coefficient’ is a myth \cite{Bendavid_PRL_2011}. He showed that, by manipulating the spatial stress profiles along an interface, we could essentially vary the static friction coefficient by over a factor of 2. These large variations were demonstrated with experiments using the same two sliding blocks. The value of the friction coefficient correlated well with the nature of the earthquakes that were observed for each experiment!
	
	{\em Why in the world are we doing this?}
	
	 After Oded graduated, Ilya Svetlizky took over this experiment. Oded’s measurements of the local stresses along the interface were very slow; each stress profile could only be measured at most 10 times a second. This was fine if we wanted to know the before and after stress profiles, but now we (I) got greedy and wanted to measure the dynamic stresses generated by these frictional ruptures as they happened.  Ilya spent 1-2 years putting together the necessary electronics and fast digitizing system to perform these experiments. When he asked me what we were looking for, I really did not have a better answer for him other than “I think that it may turn out to be interesting”. 
	 
	  Finally, the big day arrived and Ilya started taking measurements. We were rewarded with an impressive ensemble of wiggling lines.  Luckily, at about the time that we were getting our first results, my friend, Mokhtar Adda-Bedia, came to visit. While staring at the wiggles, Mokhtar, who is an accomplished theorist in the fracture business, commented that the they actually looked a lot like the angular functions corresponding to the solutions for dynamic cracks. I  had entirely missed this, although my intuition was telling me that these frictional ruptures had to, in some sense, be related to shear cracks.
	  
	   Ilya immediately pounced on this and quickly found that Mokhtar’s insight was exactly on the mark. The ‘wiggles’ perfectly corresponded to the fracture mechanics solutions for rapidly propagating shear cracks (another `wow' moment)! This result established that friction (at least the onset of friction) is a pure fracture process \cite{Svetlizky_Nature_2014}. We then went on to show that all of the quantitative predictions that were derived for shear cracks worked perfectly.  (These had never before been, actually, verified experimentally, since pure shear cracks were never really thought to exist.) Together with Elsa Bayart, then a postdoc in our group, Ilya and Elsa went on to show that (1) frictional shear cracks obeyed the predicted dynamic equation of motion \cite{Svetlizky_EOM_PRL_2017}, (2) how the wiggly lines can  be used to measure the frictional resistance (fracture energy) of frictional interfaces under a variety of conditions \cite{Bayart_JGR_2018,Bayart_lubrication_PRL_2016}. Moreover, through our measurements of the stresses along the interface, we are able to quantitatively predict if and when frictional ruptures will arrest along an interface \cite{Bayart_NatPhys_2016}. This last is equivalent to predicting the magnitude of an earthquake (before it happens) along a natural fault. While we are not yet in a position to predict real earthquakes, these experiments are valuable since they tell us what we need to find out in order to do so. This body of work, that we initiated 10 years before on a whim, has uncovered a new paradigm for understanding friction. Basically, we have shown that static friction is essentially quantitatively described by dynamic fracture.  Sometimes we get lucky. 
	   
	{\em Believe in your friends (sometimes) - but only after doing the experiment}	
	
The work described above has provided both entirely new understanding of friction and significant new insights about how earthquakes propagate, arrest and initiate. This, however, is far from the whole story of either friction or earthquake dynamics. My good friend, Yehuda Ben-zion told me, years ago, that when two bodies with {\em different} elastic properties undergo frictional sliding, the physics of this process, called bimaterial friction, can be wholly different. Since this story appealed to my contrarian nature, upon reading the relevant literature, I decided that the idea was worth checking out experimentally. This is especially true since the bimaterial case is actually the most general scenario for friction.

I should note here, something that I discovered a while ago. Many people (myself included) will not entirely believe either a theory or numerical calculations, unless there are experiments to back them up. The story of bimaterial friction is a good case in point. The ideas and initial theory are over 50 years old \cite{Weertman_1963,Weertman_1980} and simulations showed the effect to be real \cite{AmpueroBenZion2008,Shi_benzion_2006}, but experiments were still required. Why are experiments so critical? Well, nature is highly complex, and often quite fickle, with numerous unknowns. Most of the time, we don't have any idea of even what the unknowns are (they {\em are}, after all, unknowns).  Whenever we work on either an analytic theory or even simulate a system numerically, we always have to introduce significant simplifications. The only way to know if these are justified, is to ask nature what the real answer is - i.e. do a good experiment. 

Both the theory and simulations predicted that the up-down asymmetry introduced by sliding blocks composed of different materials was enough to entirely change the way that the interface `breaks'. My student, Hadar Shlomai, worked hard to test these ideas. I don't have the space (or inclination) to describe the maze of problems that she had to surmount to do a `good experiment'. These experiments have indeed revealed a new `can of worms' of exotic fracture modes along bimaterial interfaces. These are very much different than the `simple' shear cracks that populate fracture frictional interfaces, when sliding bodies are the same \cite{Shlomai_2016}. Among these are entities called `slip pulses', that behave sort of like very rapid inch-worms; highly localized pulses that only fracture the interface locally by 'lifting up'. I can say that, in this case, many of the crazy ideas that were predicted by my friends indeed are observed. Sometimes, even theorists can get it right. In addition, as always, we also found a number of quite interesting surprises. I won't describe this business any further. For this, you will have to read the papers. Some of these are being written in parallel to these ramblings.

\section{Wrapping it up}
If you have arrived at this point, you may agree with me that I have had an interesting career as an experimental physicist. While 95\% of the time in this endeavor has been pure frustration, I really don't regret much. The few and very much separated high points, the `wow' moments, are (for me) what has made this long journey worth taking. Much of what I have learned, however, is not written in any of the papers that I have published along the way. The papers describe some of the fun things that we have stumbled upon, some points along the road. They do not, however, describe the nature of the road that led us to those points. In this paper, I have tried to describe this road. While writing this account, I tried to envision a young scientist who is just starting out; a person who may be as confused as I was at his/her age. By spreading these few `bread crumbs' along the road that I have traveled (and am still traveling), I hope that they can be of some benefit to guide others. I believe that if my good friend, Yves Couder, were around, he would probably be mildly amused by this. I'm sure that he would have had many more (and more intelligent) things to say about this long and twisting road that we scientists tread. Unfortunately, he has left us way too early, so the interested reader will have to make do my own story, admittedly a poor second choice.


%
%
%
%
%
%
%
%
%
%
%
%
%
%
%
%
%


\begin{thebibliography}{10}
	
	\bibitem{Fineberg_monopoles_1985}
	J~FINEBERG.
	\newblock {Monopole pair production in compact u(1)}.
	\newblock {\em {Physics Letters B}}, {158}({2}):{135--139}, {1985}.
	
	\bibitem{Ahlers_Victor_1985}
	G~Ahlers, DS~Cannell, AND V~Steinberg.
	\newblock {Time-dependence of flow patterns near the convective threshold in a
		cylindrical container}.
	\newblock {\em {Physical Review Letters}}, {54}({13}):{1373--1376}, {1985}.
	
	\bibitem{Ahlers_berhringer_1978}
	G~Ahlers and RP~Behringer.
	\newblock {Evolution of turbulence from Rayleigh-Benard instability}.
	\newblock {\em { Physical Review Letters }}, {40}({11}):{712--716}, {1978}.
	
	\bibitem{Fineberg_vortexfront_1987}
	J~Fineberg and V~Steinberg.
	\newblock {Vortex-front propagation in Rayleigh-Benard convection}.
	\newblock {\em { Physical Review Letters }}, {58}({13}):{1332--1335}, {MAR 30}
	{1987}.
	
	\bibitem{Dee_Langer_1983}
	G~Dee and JS~Langer.
	\newblock {Propagating pattern selection}.
	\newblock {\em { Physical Review Letters }}, {50}({6}):{383--386}, {1983}.
	
	\bibitem{Moses_multistability_1987}
	E~Moses, J~Fineberg, and V~Steinberg. 
	\newblock {Multistability and confined traveling-wave patterns in a convecting
		binary mixture}.
	\newblock {\em { Physical Review A}}, {35}({6}):{2757--2760}, {MAR 15} {1987}.
	
	\bibitem{Fineberg_Moses_PRL_1988}
	J~Fineberg, E~Moses, and V~Steinberg.
	\newblock {Spatially and temporally modulated traveling-wave pattern in
		convecting binary-mixtures}.
	\newblock {\em { Physical Review Letters }}, {61}({7}):{838--841}, {AUG 15}
	{1988}.
	
	\bibitem{Marder_howthingsbreak_1996}
	M~Marder and J~Fineberg.
	\newblock {How things break}.
	\newblock {\em {Physics Today}}, {49}({9}):{24--29}, {SEP} {1996}.
	
	\bibitem{Fineberg_instabilityPRL_1991}
	J~Fineberg, SP~Gross, M~Marder, and HL~Swinney.
	\newblock {Instability in Dynamic Fracture}.
	\newblock {\em { Physical Review Letters }}, {67}({4}):{457--460}, {JUL 22}
	{1991}.
	
	\bibitem{Lathrop_Fineberg_1992}
	DP~Lathrop, J~Fineberg, and HL~Swinney.
	\newblock {Transition to shear-driven turbulence in Couette-Taylor flow}.
	\newblock {\em {Physical Review A}}, {46}({10}):{6390--6405}, {NOV 15} {1992}.
	
	\bibitem{Edwards_1994}
	WS~Edwards and S~Fauve.
	\newblock {Patterns and quasi-patterns in the Faraday experiment}.
	\newblock {\em {Journal of Fluid Mechanics}}, {278}:{123--148}, {NOV 10}
	{1994}.
	
	\bibitem{Lioubashevski_DissSolitions_1996}
	O~Lioubashevski, H~Arbell, and J~Fineberg.
	\newblock {Dissipative solitary states in driven surface waves}.
	\newblock {\em { Physical Review Letters }}, {76}({21}):{3959--3962}, {MAY 20}
	{1996}.
	
	\bibitem{Couder4}
	Emmanuel Fort, Antonin Eddi, Arezki Boudaoud, Julien Moukhtar, and Yves Couder.
	\newblock {Path-memory induced quantization of classical orbits}.
	\newblock {\em {Proceedings of the National Academy of Sciences of the uUnited
			States of America}}, {107}({41}):{17515--17520}, {OCT 12} {2010}.
	
	\bibitem{Umbanhowar_oscillons}
	PB~Umbanhowar, F~Melo, and HL~Swinney.
	\newblock {Localized excitations in a vertically vibrated granular layer}.
	\newblock {\em {Nature}}, {382}({6594}):{793--796}, {AUG 29} {1996}.
	
	\bibitem{Arbell_Oscillons_2000}
	H~Arbell and J~Fineberg.
	\newblock {Temporally harmonic oscillons in Newtonian fluids}.
	\newblock {\em { Physical Review Letters }}, {85}({4}):{756--759}, {JUL 24}
	{2000}.
	
	\bibitem{Lioubashevski_Oscillons_1999}
	O~Lioubashevski, Y~Hamiel, A~Agnon, Z~Reches, and J~Fineberg.
	\newblock {Oscillons and propagating solitary waves in a vertically vibrated
		colloidal suspension}.
	\newblock {\em { Physical Review Letters }}, {83}({16}):{3190--3193}, {OCT 18}
	{1999}.
	
	\bibitem{Arbell_PRE_2002}
	H~Arbell and J~Fineberg.
	\newblock {Pattern formation in two-frequency forced parametric waves}.
	\newblock {\em { Physical Review E}}, {65}({3, 2A}), {MAR} {2002}.
	
	\bibitem{Epstein_PRL_2008}
	T.~Epstein and J.~Fineberg.
	\newblock {Necessary conditions for mode interactions in parametrically excited
		waves}.
	\newblock {\em { Physical Review Letters }}, {100}({13}), {APR 4} {2008}.
	
	\bibitem{Shani_PRL_2010}
	Itamar Shani, Gil Cohen, and Jay Fineberg.
	\newblock {Localized Instability on the Route to Disorder in Faraday Waves}.
	\newblock {\em { Physical Review Letters }}, {104}({18}), {MAY 7} {2010}.
	
	\bibitem{Sharon_PRL_1995}
	E~Sharon, SP~Gross, and J~Fineberg.
	\newblock {Local crack branching as a mechanism for instability in dynamic
		fracture}.
	\newblock {\em {Physical Review Letters }}, {74}({25}):{5096--5099}, {JUN 19}
	{1995}.
	
	\bibitem{Sharon_energy_1996}
	E~Sharon, SP~Gross, and J~Fineberg.
	\newblock {Energy dissipation in dynamic fracture}.
	\newblock {\em { Physical Review Letters }}, {76}({12}):{2117--2120}, {MAR 18}
	{1996}.
	
	\bibitem{Morrissey_Rice_2000}
	JW~Morrissey and JR~Rice.
	\newblock {Crack front waves}.
	\newblock {\em {Journal of the Mechanics and Physics of Solids}},
	{46}({3}):{467--487}, {MAR} {1998}.
	
	\bibitem{Ramanathan_Fisher_1997}
	S~Ramanathan and DS~Fisher.
	\newblock {Dynamics and instabilities of planar tensile cracks in heterogeneous
		media}.
	\newblock {\em {Physical Review Letters }}, {79}({5}):{877--880}, {AUG 4}
	{1997}.
	
	\bibitem{Sharon_FWNature_2001}
	E~Sharon, G~Cohen, and J~Fineberg.
	\newblock {Propagating solitary waves along a rapidly moving crack front}.
	\newblock {\em {Nature}}, {410}({6824}):{68--71}, {MAR 1} {2001}.
	
	\bibitem{Sagy_Nature_2002}
	A~Sagy, Z~Reches, and J~Fineberg.
	\newblock {Dynamic fracture by large extraterrestrial impacts as the origin of
		shatter cones}.
	\newblock {\em {Nature}}, {418}({6895}):{310--313}, {JUL 18} {2002}.
	
	\bibitem{Sagy_ShatterJGR_2004}
	A~Sagy, J~Fineberg, and Z~Reches.
	\newblock {Shatter cones: Branched, rapid fractures formed by shock impact}.
	\newblock {\em {Journal of Geophysical Research-solid earth}}, {109}({B10}),
	{OCT 26} {2004}.
	
	\bibitem{Livne_PRL_2005}
	A~Livne, G~Cohen, and J~Fineberg.
	\newblock {Universality and hysteretic dynamics in rapid fracture}.
	\newblock {\em {Physical Review Letters }}, {94}({22}), {JUN 10} {2005}.
	
	\bibitem{Livne_PRL_2008}
	Ariel Livne, Eran Bouchbinder, and Jay Fineberg.
	\newblock {Breakdown of Linear Elastic Fracture Mechanics near the Tip of a
		Rapid Crack}.
	\newblock {\em { Physical Review Letters }}, {101}({26}), {DEC 31} {2008}.
	
	\bibitem{Bouchbinder_PRL_2008}
	Eran Bouchbinder, Ariel Livne, and Jay Fineberg.
	\newblock {Weakly Nonlinear Theory of Dynamic Fracture}.
	\newblock {\em { Physical Review Letters }}, {101}({26}), {DEC 31} {2008}.
	
	\bibitem{Goldman_Interia_PRL_2010}
	Tamar Goldman, Ariel Livne, and Jay Fineberg.
	\newblock {Acquisition of Inertia by a Moving Crack}.
	\newblock {\em { Physical Review Letters }}, {104}({11}), {MAR 19} {2010}.
	
	\bibitem{Marder_1991}
	M~MARDER.
	\newblock {New dynamic equation for cracks}.
	\newblock {\em { Physical Review Letters }}, {66}({19}):{2484--2487}, {MAY 13}
	{1991}.
	
	\bibitem{Livne_Science_2010}
	Ariel Livne, Eran Bouchbinder, Ilya Svetlizky, and Jay Fineberg.
	\newblock {The Near-Tip Fields of Fast Cracks}.
	\newblock {\em {Science}}, {327}({5971}):{1359--1363}, {MAR 12} {2010}.
	
	\bibitem{Bouchbinder_AnnRev_2010}
	Eran Bouchbinder, Jay Fineberg, and M.~Marder.
	\newblock {Dynamics of Simple Cracks}.
	\newblock In {Langer, JS}, editor, {\em {Annual Review of Condensed Matter
			PHYSICS, VOL 1}}, volume~{1} of {\em {Annual Review of Condensed Matter
			Physics}}, pages {371--395}. {2010}.
	
	\bibitem{Goldman_Oscillations_2012}
	Tamar Goldman, Roi Harpaz, Eran Bouchbinder, and Jay Fineberg.
	\newblock {Intrinsic Nonlinear Scale Governs Oscillations in Rapid Fracture}.
	\newblock {\em { Physical Review Letters }}, {108}({10}), {MAR 9} {2012}.
	
	\bibitem{Bouchbinder_Goldman_2014}
	Eran Bouchbinder, Tamar Goldman, and Jay Fineberg.
	\newblock {The dynamics of rapid fracture: instabilities, nonlinearities and
		length scales}.
	\newblock {\em {Reports on Progress in Physics}}, {77}({4}), {APR} {2014}.
	
	\bibitem{Karma_bouchbinder_NatPhys_2017}
	Chih-Hung Chen, Eran Bouchbinder, and Alain Karma.
	\newblock {Instability in dynamic fracture and the failure of the classical
		theory of cracks}.
	\newblock {\em {Nature Physics}}, {13}({12}):{1186+}, {DEC} {2017}.
	
	\bibitem{Goldman_microbranchin_PRL_2015}
	Tamar~Goldman Boue, Gil Cohen, and Jay Fineberg.
	\newblock {Origin of the Microbranching Instability in Rapid Cracks}.
	\newblock {\em { Physical Review Letters }}, {114}({5}), {FEB 2} {2015}.
	
	\bibitem{Kolvin_PRL_2015}
	Itamar Kolvin, Gil Cohen, and Jay Fineberg.
	\newblock {Crack Front Dynamics: The Interplay of Singular Geometry and Crack
		Instabilities}.
	\newblock {\em { Physical Review Letters }}, {114}({17}), {MAY 1} {2015}.
	
	\bibitem{Kolvin_Mokhtar_PRL_2017}
	Itamar Kolvin, Jay Fineberg, and Mokhtar Adda-Bedia.
	\newblock {Nonlinear Focusing in Dynamic Crack Fronts and the Microbranching
		Transition}.
	\newblock {\em { Physical Review Letters }}, {119}({21}), {NOV 21} {2017}.
	
	\bibitem{Kolvin_Nat_mat_2018}
	Itamar Kolvin, Gil Cohen, and Jay Fineberg.
	\newblock {Topological defects govern crack front motion and facet formation on
		broken surfaces}.
	\newblock {\em {Nature Materials}}, {17}({2}):{140+}, {FEB} {2018}.
	
	\bibitem{Gong_2003}
	JP~Gong, Y~Katsuyama, T~Kurokawa, and Y~Osada.
	\newblock {Double-network hydrogels with extremely high mechanical strength}.
	\newblock {\em {Advanced Materials}}, {15}({14}):{1155+}, {JUL 17} {2003}.
	
	\bibitem{Kolvin_Kolinsky_PRL_2018}
	Itamar Kolvin, John~M. Kolinski, Jian~Ping Gong, and Jay Fineberg.
	\newblock {How Supertough Gels Break}.
	\newblock {\em { Physical Review Letters }}, {121}({13}), {SEP 26} {2018}.
	
	\bibitem{Rubinstein_Nature_2004}
	SM~Rubinstein, G~Cohen, and J~Fineberg.
	\newblock {Detachment fronts and the onset of dynamic friction}.
	\newblock {\em {Nature}}, {430}({7003}):{1005--1009}, {AUG 26} {2004}.
	
	\bibitem{Bendavid_Science_2010}
	Oded Ben-David, Gil Cohen, and Jay Fineberg.
	\newblock {The Dynamics of the Onset of Frictional Slip}.
	\newblock {\em {Science}}, {330}({6001}):{211--214}, {OCT 8} {2010}.
	
	\bibitem{Bendavid_PRL_2011}
	Oded Ben-David and Jay Fineberg.
	\newblock {Static Friction Coefficient Is Not a Material Constant}.
	\newblock {\em { Physical Review Letters }}, {106}({25}), {JUN 20} {2011}.
	
	\bibitem{Svetlizky_Nature_2014}
	Ilya Svetlizky and Jay Fineberg.
	\newblock {Classical shear cracks drive the onset of dry frictional motion}.
	\newblock {\em {Nature}}, {509}({7499}):{205+}, {MAY 8} {2014}.
	
	\bibitem{Svetlizky_EOM_PRL_2017}
	Ilya Svetlizky, David~S. Kammer, Elsa Bayart, Gil Cohen, and Jay Fineberg.
	\newblock {Brittle Fracture Theory Predicts the Equation of Motion of
		Frictional Rupture Fronts}.
	\newblock {\em { Physical Review Letters }}, {118}({12}), {MAR 21} {2017}.
	
	\bibitem{Bayart_JGR_2018}
	E.~Bayart, I.~Svetlizky, and J.~Fineberg.
	\newblock {Rupture Dynamics of Heterogeneous Frictional Interfaces}.
	\newblock {\em {Journal of Geophysical Research-solid earth}},
	{123}({5}):{3828--3848}, {MAY} {2018}.
	
	\bibitem{Bayart_lubrication_PRL_2016}
	E.~Bayart, I.~Svetlizky, and J.~Fineberg.
	\newblock {Slippery but Tough: The Rapid Fracture of Lubricated Frictional
		Interfaces}.
	\newblock {\em { Physical Review Letters }}, {116}({19}), {MAY 10} {2016}.
	
	\bibitem{Bayart_NatPhys_2016}
	Elsa Bayart, Ilya Svetlizky, and Jay Fineberg.
	\newblock {Fracture mechanics determine the lengths of interface ruptures that
		mediate frictional motion}.
	\newblock {\em {Nature Physics}}, {12}({2}):{166+}, {FEB} {2016}.
	
	\bibitem{Weertman_1963}
	J~WEERTMAN.
	\newblock {Unstable slippage across a fault that separates elastic media of
		different elastic constants}.
	\newblock {\em {Journal of the Mechanics and Physics of Solids}},
	{11}({NB3}):{197--204}, {1963}.
	
	\bibitem{Weertman_1980}
	J~WEERTMAN.
	\newblock {Unstable slippage across a fault that separates elastic media of
		different elastic-constants}.
	\newblock {\em { Journal of Geophysical Research }}, {85}({NB3}):{1455--1461},
	{1980}.
	
	\bibitem{AmpueroBenZion2008}
	J.~P. Ampuero and Y.~Ben-Zion.
	\newblock Cracks, pulses and macroscopic asymmetry of dynamic rupture on a
	bimaterial interface with velocity-weakening friction.
	\newblock {\em Geophysical Journal International}, 173(2):674--692, MAY 2008.
	
	\bibitem{Shi_benzion_2006}
	ZQ~Shi and Y~Ben-Zion.
	\newblock {Dynamic rupture on a bimaterial interface governed by slip-weakening
		friction}.
	\newblock {\em { Geophysical Journal International }}, {165}({2}):{469--484},
	{MAY} {2006}.
	
	\bibitem{Shlomai_2016}
	Hadar Shlomai and Jay Fineberg.
	\newblock {The structure of slip-pulses and supershear ruptures driving slip in
		bimaterial friction}.
	\newblock {\em {Nature Communications}}, {7}, {JUN} {2016}.
	
\end{thebibliography}


\end{document}